\documentclass[prb,preprint]{revtex4}
\usepackage{amsmath}
\usepackage{amsfonts}
\usepackage{graphicx}

\begin{document}

\title{
Interpretation of Light Scattering Spectra in Terms of Particle Displacements
 }

\author{George D. J. Phillies}
\email{phillies@wpi.edu}

\affiliation{Department of Physics, Worcester Polytechnic
Institute,Worcester, MA 01609}

\begin{abstract}

Quasi-elastic light scattering spectroscopy is regularly used to examine the dynamics of dilute solutions of diffusing mesoscopic probe particles in fluids.  For probes in a simple liquid, the light scattering spectrum is a simple exponential; the field correlation function $g^{(1)}(q,\tau)$ of the scattering particles is related to their mean-square displacements $\overline{X^{2}} \equiv \langle (\Delta x(\tau))^{2} \rangle$ during $\tau$ via $g^{(1)}(q,\tau) = \exp( - \frac{1}{2} q^{2} \overline{X^{2}})$.  However, demonstrations of this expression refer only to identical Brownian particles in simple liquids, and show that if the form is correct \emph{then it is also true} for all $\tau$ that $g^{(1)}(q,\tau) = \exp( - \Gamma \tau)$, a pure exponential in $\tau$.  In general,  $g^{(1)}(q,\tau)$ is not a single exponential in time.   A correct general form for $g^{(1)}(q,\tau)$ in terms of the $\overline{X^{2n}}$, replacing the incorrect $\exp( - \frac{1}{2} q^{2} \overline{X^{2}})$, is obtained.  A simple experimental diagnostic determining when the field correlation function gives the mean-square displacement is identified, namely $g^{(1)}(q,\tau)$ only reveals $\overline{X^{2}}$ if  $g^{(1)}(q,\tau)$ is a single exponential in $\tau$.  Contrariwise, if $g^{(1)}(q,\tau)$ is not a single exponential, then $g^{(1)}(q,\tau)$ depends not only on $\overline{X^{2}}$ but on all higher moments $\overline{X^{2n}}$.  Corrections to the crude approximation $g^{(1)}(q,\tau) = \exp( - \frac{1}{2} q^{2} \overline{X^{2}})$ closely resemble the higher spectral cumulants from a  cumulant expansion of 
$g^{(1)}(q,\tau)$.  
\end{abstract}

\maketitle

\section{Introduction}

Recently, there has been interest in describing light scattering spectra in terms of mean-square displacements of the scattering particles.  This description of scattering spectra is readily traced back to Berne and Pecora's book\cite{berne1976a}, where it is shown that the field correlation function $g^{(1)}(q,\tau)$ from a quasielastic light scattering spectrum   of \emph{dilute, true Brownian particles} is determined by their mean-square displacements $ \langle (\Delta \mathbf{R})^{2} \rangle$ through various times $\tau$, namely
\begin{equation}
    g^{(1)}(q,\tau) = \exp(- \langle (\Delta \mathbf{R})^{2} \rangle q^{2} /6),
    \label{eq:g1deltar2}
\end{equation}  
where $q$ is the scattering vector. The calculation of Berne and Pecora was meant to describe experiments on, for example, dilute solutions of monodisperse polystyrene latex spheres in pure water.

Prior to Berne and Pecora's work, Hallett and students\cite{hallett1974a,hallett1976a} used quasielastic light scattering spectroscopy (QELSS) to study diffusion of dilute scattering spheres, used as optical probes, through complex fluids, namely hyaluronic acid and dextran solutions.  The sphere diffusion coefficient was combined with the Stokes-Einstein equation
\begin{equation}
     D = \frac{k_{B}T}{6 \pi \eta R}
     \label{eq:SEeq}
\end{equation}
to infer a viscosity, which was compared with macroscopic measurements.  Here $k_{B}$, $R$, and $T$ are Boltzmann's constant, the probe radius, and the absolute temperature. This optical probe diffusion technique was a natural extension of studies by Laurent, et al.\cite{laurent1963a} of sedimentation of proteins and polystyrene spheres through dextran and other polymer solutions. 

The optical probe method has since been extensively been applied to colloid\cite{pusey1982a}, surfactant\cite{phillies1993a}, and polymer solution systems\cite{russo1993a,lin1982a}.  Techniques for avoiding polymer binding by probes have long been understood\cite{russo1989a}.  Extensive information on the effects of polymer molecular weight and concentration on probe diffusion has been obtained\cite{yang1988a,furukawa1991a}; changes in probe diffusion in polymer solutions, attendant to changes in solvent quality, have been determined\cite{clomenil1993a}. Rigid and flexible probes have been compared\cite{cao1997a}. The diffusion of hard-sphere probes has been contrasted with the diffusion of flexible-polymer probes through the same polymer solutions, solutions being more effective at retarding chain probes than at retarding the motion of spheres through the same solutions.\cite{pu1989a}  Nonexponential spectra have been analyzed by cumulant expansions, Laplace inversion, and multi-mode fitting.  The diffusion of polymer probes through polymeric matrix solutions, whether observed via QELSS or other techniques notably fluorescence recovery after photobleaching and pulsed-field-gradient NMR, is the topic of polymer self and tracer diffusion, whose literature has recently been systematically reviewed\cite{phillies2004a}.
                                                                
Since Hallet's invention of the optical probe diffusion method, additional experimental methods have become available. QELSS examines single-scattered photons.  With increasing probe concentration multiple scattering becomes more important, but may be suppressed with a two-laser-beam two-detector homodyne coincidence spectroscopy (HCS) apparatus\cite{phillies1981a}. Popescu, et al.,\cite{popescu2002a} have shown how short-coherence-length light can be used to restrict significant scattering to small volumes from which only single scattering contributes to the spectrum.  Alternatively, diffusing wave spectroscopy (DWS) takes full advantage of the properties of light scattered by a profoundly multiply-scattering fluid\cite{pine1988a}.  All techniques can reach the same time scales, because all techniques use the same photodetectors and correlators, but because DWS responds to the joint motion of many particles it includes relaxations due to motions of single particles that are too small to lead to appreciable relaxations in QELSS spectra. Note that in the optical probe community 'dynamic light scattering' (DLS) refers exclusively to QELSS, while the microrheology community uses 'dynamic light scattering' as including DWS, so each reference to DLS must be interpreted against experimental details.

Under the cognomen {\em microrheology}, an alternative school for the study of optical probe diffusion has recently appeared\cite{mason1995a}. Suppose that the mean-square particle displacement of particles in an \emph{arbitrary} liquid were available.  This supposition was not advanced by Ref.\ \onlinecite{berne1976a} in their presentation of eq \ref{eq:g1deltar2}.  A non-exponential QELSS or other light-scattering spectrum $S(q, \tau)$ could then formally be said to define a time-dependent diffusion coefficient, via
\begin{equation}
    g^{(1)}_{P}(q,\tau) = \exp(- \tau q^{2} D(\tau)),
    \label{eq:dq2t}
\end{equation}
where $q$ is the magnitude of the experimental scattering vector and $\tau$ is the delay time.
Laplace transform of $D(\tau) \tau$, interpreted as $\overline{\Delta R^{2}}$, and application of a generalized Stokes-Einstein equation in which $D$ and $\eta$ are generalized to frequency-dependent forms $\tilde{D}(\omega)$ and $\eta(\omega)$, might then be proposed as a path to measuring a frequency-dependent complex modulus $\tilde{G}(\omega)$.

This paper treats the field correlation function determined with QELSS. The interpretation of experiments using DWS is not considered.  Sections II and III of this paper recall at minimum length the historical approaches that do lead to eq \ref{eq:g1deltar2} under its assumed conditions, and that might be said to lead to eq  \ref{eq:dq2t}. Section II covers Berne and Pecora's treatment\cite{berne1976a} of the field correlation function for dilute Brownian particles, while Section III covers Doob's First Theorem\cite{doob1942a} for correlation functions of stationary, jointly Gaussian-random, Markoff processes.  Section IV obtains the actual relationship between the scattering spectrum and the mean-square and higher moments of the distribution of particle displacements.  \emph{If} the light-scattering spectrum of a suspension of dilute particles is determined by the mean-square particle displacement, then it is \emph{necessarily} also true that the light scattering spectrum is a simple exponential. Conversely, if the light scattering spectrum is not a simple exponential, for example if it has a secondary slow mode or if it decays as a stretched exponential in time at large time, then the spectrum is not determined by the mean-square particle displacements through different times.  Section V analyzes in detail a specific model system, thereby clarifying issues related to low-$q$ limits and spectral cumulants.  A Discussion closes the paper.

\section{Light Scattering Spectrum of Ideal Brownian Particles}

This Section considers the light scattering spectrum of a dilute solution of identical true Brownian particles, showing that in this special case the QELSS spectrum does determine the mean-square particle displacement during different intervals. In a QELSS experiment, an equilibrium solution or suspension is illuminated by a narrow pencil of coherent light.  The light scattered by the solution through a fixed angle is collected.  The time-dependent intensity $I(t)$ of the collected light is measured.  In time domain, the QELSS spectrum $S(q,t)$ is obtained as the time correlation function 
\begin{equation}
    S(q, \tau) = \langle I(t) I(t+\tau) \rangle.
    \label{eq:sqt}
\end{equation}
Here $q$ is the magnitude of the scattering vector, $\tau$ is the delay time, and the brackets $\langle \cdots \rangle$ denote a time average over $t$.  Under normal experimental conditions, the time dependence of the measured light scattering spectrum is determined by the field correlation function $g^{(1)}(q, \tau)$, namely
\begin{equation}
    S(q, \tau) = A [g^{(1)}(q, \tau)]^{2} + B.
    \label{eq:sg1qt}
\end{equation}
Here $B$ is the time-independent baseline and $A$, the amplitude, is a partially apparatus-dependent constant.  Several normalizations are in common use for $A$ and $B$. The $S(q,t)$ that is directly computed by a real digital correlator is a series of large integers, each of which is a sum of large numbers of terms $n_{j} n_{j+\tau}$, $n_{j}$ and $n_{j+\tau}$ being the integer numbers of photons counted during short time intervals centered on absolute times $j$ and $j+\tau$.

$g^{(1)}(q,\tau)$ is in turn determined by the positions ${\bf r}_{i}(t)$ of each of the $N$ scattering particles at various times. Objects that do not scatter light may affect the spectrum, but only indirectly, by modifying the dynamics of the scattering particles. In the limit that the concentration of scattering particles is small, the same-time positions of pairs of scatterers are uncorrelated, and the field correlation function reduces to
\begin{equation}
     g^{(1)}(q,\tau) = \left\langle \sum_{i=1}^{N} \alpha_{i}^{2} \exp[i \mathbf{q} \cdot \Delta
\mathbf{R}_{i}(\tau)] \right\rangle.
     \label{eq:g1defP}
\end{equation}
Here $g^{(1)}(q, \tau)$ is the \emph{incoherent structure factor}, termed \emph{incoherent} because it does not include terms involving distinct pairs of scatterers.  The term $\alpha_{i}^{2}$ is a normalized scattering cross-section for particle $i$. If all particles scatter equally, $\alpha_{i}^{2}$ may be absorbed into $A$.  The displacement of particle $i$ during $\tau$ is $\Delta \mathbf{R}_{i}(\tau) = \mathbf{r}_{i}(t+\tau) - \mathbf{r}_{i}(t))$.
Each displacement is an integral of the particle velocity $\mathbf{v}_{i}(s)$ over times $(t, t+\tau)$,
\begin{equation}
    \Delta  \mathbf{R}_{i}(\tau) = \int_{t}^{t+\tau} ds \ \mathbf{v}_{i}(s).
    \label{eq:drv}
\end{equation}

Further analysis requires a physical description of the particle motions.  In the simplest idealization, noninteracting probe particles in solution independently perform ideal Brownian motion as described by the Langevin equation.  This special case is treated by Berne and Pecora\cite{berne1976a}, who show from the Central Limit Theorem that particle displacements $\Delta \mathbf{R}$, over times $\tau$ much longer than the velocity autocorrelation function's relaxation time, have for ideal Brownian particles a probability distribution  
\begin{equation}
     G_{s}(\Delta \mathbf{R}, \tau) = \left[ \frac{2 \pi}{3} \langle (\Delta \mathbf{R})^{2} \rangle\right]^{-3/2} \exp\left[ - 3 (\Delta R)^2/ 2
\langle (\Delta \mathbf{R})^{2} \rangle\right].
     \label{eq:deltaR}
\end{equation}

For Brownian particles, the average displacement over $\tau$ is
\begin{equation}
   \langle (\Delta \mathbf{R})^{2} \rangle = 6 D \tau,
   \label{eq:deltaRval}
\end{equation}
where $D$ is the single-particle diffusion coefficient of the Stokes-Einstein equation.

The incoherent structure factor is then 
\begin{equation}
     g^{(1)}(q,\tau) = \int d  \Delta \mathbf{R} \   G_{s}(\Delta \mathbf{R}, \tau)  \sum_{i=1}^{N} \exp[i \mathbf{q} \cdot \Delta
\mathbf{R}_{i}(\tau)]     \sim \exp(-  D q^{2} \tau)
   \label{eq:g1calc2}
\end{equation}
An equivalent form is eq \ref{eq:g1deltar2}, as seen in Berne and Pecora\cite{berne1976a}.  Eq \ref{eq:g1deltar2} is sometimes cited as arising from Pusey and Tough's treatment\cite{pecora1985a} of interacting particle systems, but section 4.3.2 of Ref \onlinecite{pecora1985a} scrupulously footnotes its derivation of eq \ref{eq:g1deltar2} as coming from Ref \onlinecite{berne1976a}.

For simple Brownian particles, the natural logarithm of $g^{(1)}(q, \tau)$ is proportional to the mean-square displacement of a particle during $\tau$, and is linear in $\tau$, so the slope  $d \log(g^{(1)}(q, \tau)/ d \tau$ can be used to determine $D$.  In practice,  $g^{(1)}$ is never quite a perfect single exponential, and recourse must be had to a more elaborate analysis, such as the cumulant expansion\cite{cumulants}.

As emphasized by Berne and Pecora\cite{berne1976a}, their calculation of $g^{(1)}(q, \tau)$ refers to noninteracting particles performing ideal Brownian motion as governed by the simple Langevin equation, in which the external random force has a correlation time of zero.  This Brownian motion idealization is not applicable to any physical system.  In particular, the calculation does not describe noninteracting spheres diffusing in a small-molecule Newtonian solvent.  In a real Newtonian liquid, the hydrodynamic wake around a moving particle creates a memory effect.  The 'random' force on a probe particle thereby gains a long-time memory term, and does not have a correlation time of zero.  The memory term gives each sphere a velocity autocorrelation function that decays at long time as $\tau^{-3/2}$, giving $g^{(1)}(q,\tau)$ a tail that does not decay exponentially at long times.  Paul and Pusey\cite{pusey1981a} have seen this tail experimentally in the Brownian motion of large polystyrene latex spheres. 

\section{Light Scattering Spectrum on the Basis of Doob's First Theorem}

An alternative calculation of the light scattering spectrum follows from Doob's treatment\cite{doob1942a} of Brownian motion and Ornstein-Uhlenbeck processes.  Doob's actual interest was a particle performing one-dimensional Brownian motion.  The particle had a position $x(t)$, dependent on time $t$, and a nominal velocity $u(t) = d x(t)/dt$.  Credible mathematical properties of $u(t)$ were shown by Doob to lead automatically to the temporal autocorrelation function $\langle u(t) u(t+\tau) \rangle$.   We first recount the key points of Doob's calculation and then show how it predicts spectra.

Restated in modern terms, Doob's First Theorem (Theorem 1.1 of ref.\ \onlinecite{doob1942a}) provides

Let $u(t)$ (with $-\infty < t < \infty$ allowed) be a family of random variables, with a single parameter $t$ labelling the members of the family, determining a stochastic process with the following properties:

(1) The process is temporally homogeneous, i.e., the distribution function for $u(t)$ is independent of $t$.

(2) If $s$ and $s'$ are two arbitrary, distinct values of $t$, then $u(s)$ and $u(s')$ have a non-singular bivariate Gaussian distribution.

(3) The process governing the evolution of $u(t)$ is a Markoff process.

Define $m$ and $\sigma_{0}^{2}$ by  $m = \langle u(t) \rangle$, $\sigma_{0}^{2} = \langle (u(t) - m)^{2} \rangle$, and $\rho(t) = \langle (u(s)-m)(u(s+t)-m) \rangle$, with $\langle \ldots \rangle$ denoting the expectation value.  Because the process is stationary, only one value for $m$ and one value for $\sigma_{0}$ exist.  Then the stochastic process is of one of two types:

(A) If $s \neq s'$, then $u(s)$ and $u(s')$ are independent, Gaussianly-distributed random variables, \emph{or}

(B) There is a constant $\beta > 0$ such that if times $s_{i}$ satisfy $s_{1} < s_{2} < \ldots < s_{n}$, then $u(s_{1}), u(s_{2}), \ldots u(s_{n})$, have an $n$-variable joint Gaussian distribution, with $\langle (u(s_{i}) -m)(u(s_{j} - m) \rangle = \sigma_{0}^{2} \exp(- \beta \mid s_{i} -s_{j} \mid)$.

To demonstrate this result, Doob observes: We may without changing anything important rescale any $u(s)$ so $m=0$ and $\sigma_{0}=1$.  From (2), the conditional distribution for $u(t)$, given a specified $u(s)$, is then
\begin{equation}
   \frac{1}{(2 \pi)^{1/2} (1 - \rho^{2})^{1/2}} \exp\left( \frac{[u(t)-\rho u(s)]^{2}}{2(1-\rho^{2})} \right)
   \label{eq:doobbivariate}
\end{equation}
with $\rho = \rho(t-s)$.  Consider an ordered series of times $t_{1} < t_{2} <\ldots < t_{n}$, and define $u_{i} = u(t_{i})$, for $i \in [1, n]$.  From (2), any pair $u_{i}, u_{j}$ with $i \neq j$ have the joint distribution of eq \ref{eq:doobbivariate}; stationarity ensures that $\rho$ only depends on the time difference $\mid t_{i}-t_{j} \mid$.
Eq \ref{eq:doobbivariate} only gives the two-time distribution function for the $u_{i}$.  From the Markoff property (3), the $n$-time distribution function is determined, because each $u_{i}$ is sensitive only to the value of the most immediately previous $u_{i-1}$, so the $u_{i}$ have an $n$-variable joint Gaussian distribution, namely
\begin{equation}
   \frac{1}{(2 \pi)^{n/2} \prod_{i=1}^{n-1}(1 - \rho_{j}^{2})^{1/2}} \exp\left(- u_{1}^{2}/2- \frac{1}{2} \sum_{i=1}^{n-1} \frac{[u_{j+1}-\rho_{j} u_{j}]^{2}}{(1-\rho_{j}^{2})} \right)
   \label{eq:doobnvariate}
\end{equation}
for $\rho_{j} = \rho(t_{j+1}-t_{j})$.

Choosing $n=3$, direct integration of eq \ref{eq:doobnvariate} shows 
\begin{equation}
    \rho(t_{3}-t_{1}) = \rho(t_{3}-t_{2}) \rho(t_{2}-t_{1}).
    \label{eq:doobrho}
\end{equation} 
It follows that $\rho$ is an even function, that $\rho(t) \leq 1$, and therefore either $\rho \equiv 0$ (outcome (A)) or
\begin{equation}
     \rho(t) = \exp(- \beta \mid t\mid )
     \label{eq:doobexponential}
\end{equation} 
for some positive $\beta$.  Eq \ref{eq:doobexponential} is clearly consistent with eq \ref{eq:doobrho}; Doob implies but does not explicitly prove uniqueness.

It is important to emphasize that it is the Markoff property of $u(s)$ that leads to eqs \ref{eq:doobnvariate} and \ref{eq:doobrho}.  The joint Gaussian property of eq \ref{eq:doobbivariate} in and of itself does not imply that the $u_{i}$ have an $n$-variable joint gaussian distribution function; it only implies that any pair of the $u_{i}$ have a two-variable joint distribution function.  In a non-Markoff system, $u_{n}$ is simultaneously sensitive to all $u_{j}$, $j < n$, not only to the most recent past $u_{j}$ to be specified.  The derivation in Section II did invoke the Markoff property of $u(s)$, though not explicitly:  Equation \ref{eq:deltaR} covertly but necessarily follows from the statement that a displacement over a large time can be described as a sum of a series of \emph{independent} displacements, each  over a shorter time. The statement that successive Brownian displacements are independent is the Markoff assumption, describing the differential pieces of each Brownian displacement; it is not the bivariate Gaussian distribution assumption.

Doob's derivation of his results depends only on mathematical properties (1)-(3), not on the physical nature of $u(s)$.  The velocity $u(s)$ of a Brownian particle does have these three properties, so the Brownian velocity does satisfy (B), but that occurs because the Brownian velocity $u(s)$ is stationary, has a two-time joint Gaussian distribution, and is a Markoff process, not because $u(s)$ represents the rate at which an ideal Brownian particle is changing its position. Correspondingly, any other variable, that has the same mathematical properties that $u(s)$ has, will also satisfy Doob's theorem.

We now consider for which systems the time dependence of the incoherent structure factor can be obtained with Doob's First Theorem. The key step is to identify an appropriate variable to replace $u$.  An interesting choice of variable is the $q^{\rm th}$ spatial Fourier component of the scatterer concentration
\begin{equation}
     a_{q}(s) = \sum_{i=1}^{N} \exp\left(- {\bf q} \cdot {\bf r}_{i}(s)\right), 
     \label{eq:aqdefinition}
\end{equation}
the incoherent structure factor being proportional to $\langle a_{q}(s) a^{*}_{q}(s+\tau) \rangle$.  Does $a_{q}(s)$ satisfy conditions (1)-(3) of the theorem?  

(1) The $a_{q}(s)$, labelled by the one parameter $s$, are measured on an equilibrium system, so they represent a stationary process.  

(2) At each time $s$, $a_{q}(s)$ is the sum of a large number $N$ of independent random functions $\exp\left(- {\bf q} \cdot {\bf r}_{i}(s)\right)$.  The individual random functions are identically distributed, so by the Central Limit Theorem their sum has a Gaussian random distribution. Furthermore, the change in $a_{q}(s)$ between two times is determined by
\begin{equation}
    \Delta a_{q}(t-s) \equiv a_{q}(t) - a_{q}(s) = \sum_{i=1}^{N}  (\exp[-i {\bf q}\cdot {\bf r}_{i}(t)]) (1 -\exp[-i {\bf q}\cdot ({\bf r}_{i}(s) - {\bf r}_{i}(t)  )]).
\end{equation}
The final form is a sum of $N$ terms.  Motions of different particles are independent, so the $N$ terms are independent from each other.  Each term is determined by the particle displacement ${\bf r}_{i}(s) - {\bf r}_{i}(t)$ between times $s$ and $t$.  If all particles are the same, the distribution of displacements is the same for every particle, and the same for all particle initial or final locations, so $\Delta a_{q}(t-s)$ is the sum of a large number of independently-distributed random variables.  From the Central Limit Theorem, $\Delta a_{q}(t-s)$ therefore has a Gaussian random distribution.  The convolution of two Gaussian distributions is a Gaussian, so $a_{q}(t) = a_{q}(s) + \Delta a_{q}(t-s)$ also has a Gaussian distribution, and $a_{q}(s)$ and $a_{q}(t)$ have a joint Gaussian distribution.

(3) If an individual particle is performing Brownian motion, and all particles are the same, information on the position ${\bf r}_{i}(t_{j})$ is given by the position at the most recent previous time $t_{j-1}$.  If the particles have no memory, once ${\bf r}_{i}(t_{j-1})$ is given, information on the particle's position at earlier times ${\bf r}_{i}(t_{j-\ell})$, $\ell >1$, gives no further information on how the particle moves between $t_{j-1}$ and $t_{j}$.  That is, for Brownian particles the ${\bf r}_{i}(t)$ and each of their functions is governed by a Markoff process. This Markoff behavior does not follow from the Central Limit Theorem rationale used to defend (2), because the displacement of each particle between $t_{1}$ and $t_{2}$ could be correlated with the displacement of the same particle between $t_{2}$ and $t_{3}$.

Under these rather restrictive conditions on particle motion, $a_{q}(t)$ has the same mathematical properties as the $u(t)$ examined by Doob..  In particular $a_{q}(t_{1})$, $a_{q}(t_{2})$, and $a_{q}(t_{3})$ have a three-variable joint Gaussian distribution.  The autocorrelation function of $a_{q}(t)$ is therefore the same as the autocorrelation function of $u(t)$, leading to an exponentially-decaying form
\begin{equation}
      \langle a_{q}(s) a^{*}_{q}(t) \rangle = \sigma_{0}^{2} \exp (- \beta \mid t-s \mid).
      \label{eq:aq2}
\end{equation}

The incoherent structure factor is almost never a pure exponential.  Even the slight polydispersity of the spheres in a polystyrene latex preparation leads to a measurable non-exponentiality of the incoherent structure factor.  Where does this non-exponentiality arise, relative to the prediction of Doob's theorem?  The key issue is that if the system is polydisperse, then all particles are not the same; some are more mobile than others, and the mobile particles continue to be more mobile as time passes.   In a series of times $t_{1}$, $t_{2}$, $t_{3}$, the difference $a_{q}(t_{2}) - a_{q}(t_{1})$ gives information about the likelihood of changes $a_{q}(t_{3}) - a_{q}(t_{2})$.  For example, a particularly small change in $a_{q}$ between times $t_{1}$ and $t_{2}$ suggests that the fluctuation in $a_{q}$ initially involved an unusually large number of larger, less mobile particles, a condition that will tend to persist between times $t_{2}$ and $t_{3}$, so if the change in $a_{q}(t)$ between $t_{1}$ and $t_{2}$ is slow, the change in $a_{q}$ between $t_{2}$ and $t_{3}$ is also more likely than usual to be slow.  In consequence, for a polydisperse system $a_{q}(t)$ is not governed by a Markoff process.  Even though its $a_{q}(t)$ at every pair of times has a joint Gaussian distribution, if $a_{q}(t)$ is not governed by a Markoff process $a_{q}(t)$ at a trio of times is not described by a three-variable joint Gaussian distribution.  Without the Markoff condition, an equation like \ref{eq:doobrho} but with $a_{q}(t)$ replacing $u(t)$ is invalid. Doob's analysis thus does not predict that $\langle a_{q}(s) a^{*}_{q}(t) \rangle $ of a polydisperse suspension is a decaying exponential.

As a further example of a non-Markoff scattering experiment, consider light scattering electrophoresis.  In a series of times $t_{1}$, $t_{2}$, $t_{3}$, the particle displacement between $t_{1}$ and $t_{2}$ gives information about the expected displacement between $t_{2}$ and $t_{3}$.  In consequence, the results here do not lead to the incorrect prediction that the time correlation function in a light scattering electrophoresis experiment is a decaying pure exponential.

The above refers to non-exponential spectra arising from probe polydispersity.  However, any other effect that leads to a particle motion with memory, such as viscoelastic effects giving long-time correlations to the random force on each particle, has the same consequences:   Memory effects are present, so three $a_{q}(t)$ do not necessarily have a trivariate joint Gaussian distribution. The spectrum is not a decaying single exponential. The non-Gaussian behavior deduced from this consideration lies in the $a_{q}(t)$, not directly in the distribution of particle displacements.  However, if successive displacements of every particle had the same multivariate joint Gaussian distribution function, then three $a_{q}(t)$ would also have a trivariate joint Gaussian distribution, eq \ref{eq:doobrho} would follow, and the spectrum would be a pure decaying exponential.  Therefore, if the incoherent structure factor is nonexponential, not only trios of $a_{q}(t)$ but also trios of successive particle positions do not have three-variable joint-Gaussian distribution. 

\section{Light Scattering Spectrum by Direct Calculation}

Superficially, it would appear that the light scattering spectrum could be calculated via a Taylor series expansion of eq \ref{eq:g1defP}, namely

\begin{equation}
     g^{(1)}(q,\tau) = \sum_{i=1}^{N} \left\langle \sum_{n=0}^{\infty}  \frac{(i \mathbf{q} \cdot \Delta
\mathbf{R}_{i}(\tau))^{n} }{n!} \right\rangle
     \label{eq:g1defP2}
\end{equation}
$n$ being the Taylor expansion index, with the $N$ particles potentially not all being the same.  Inversion symmetry causes all terms odd in $\Delta \mathbf{R}_{i}(\tau)$ to average to zero. For identical Brownian particles that follow  eq \ref{eq:deltaR} and $m$ a positive integer
\begin{equation}
  \left \langle (\mathbf{q} \cdot \Delta \mathbf{R}_{i}(\tau))^{2 m} \right \rangle =  q^{2m} 2^{m} (D t)^{m} (2m-1) (2m-3) \ldots (1).
   \label{eq:gaussianterms}
\end{equation}
On substituting eq \ref{eq:gaussianterms} into eq \ref{eq:g1defP2}, the factors $(2m-1)(2m-3) \ldots$ cancel the odd factors of $n!$, while the term $2^{m}$ is cancelled by the factors of 2 in the even factors of $n!$, leading to
\begin{equation}
     g^{(1)}(q,\tau) = \sum_{i=1}^{N} \sum_{n=0}^{\infty}  \frac{(- D q^{2} t)^{n}}{n!}\equiv \sum_{i=1}^{N} \exp(- q^{2} \overline{X^{2}} /2),
      \label{eq:g1defP2a}
\end{equation}
where $\overline{X^{2}} =\langle (\Delta x_{i}(\tau))^{2} \rangle$ and where $\Delta x_{i}(\tau) = \hat{q} \cdot \Delta {\mathbf{R}}_{i}(\tau)$.  For dilute Brownian particles, the Taylor expansion approach recovers the results of the previous sections.

What if the diffusing particles do not execute simple Brownian motion?  The difference between the actual  $g^{(1)}(q,\tau)$ and its Markoff-process component is seen by the factorization
\begin{equation}
      g^{(1)}(q,\tau) = \exp(- q^{2} \overline{X^{2}} /2)  \left\langle \exp
      \left( i \mathbf{q} \cdot \Delta \mathbf{R}(\tau))+  q^{2} \overline{X^{2}} /2      \right) \right\rangle.
            \label{eq:g1Pdifference}
\end{equation}
Within the average, expanding separately the exponentials in $\Delta \mathbf{R}(\tau)$ and $\overline{X^{2}}$, sorting terms by their order in $q$, and re-exponentiating, one finds the approximation
\begin{equation}
      g^{(1)}(q,\tau) = \exp\left[-\frac{q^{2}}{2} \overline{X^{2}}\right] \{1 + \frac{q^{4}}{24} (\langle (\Delta x(\tau))^{4} \rangle -  3 (\overline{X^{2}})^{2} )      - {\cal O}(q^{6}) + \ldots \}   
 \label{eq:g1Pdifference2}
\end{equation}
Define $\overline{X^{n}} =  \langle (\Delta x(\tau))^{n} \rangle$.  The terms $1 + a q^{4}$ are reorganized by multiplying by unity in the form $\exp(a q^{4}) \exp(-a q^{4})$, factoring an $\exp(+ a q^{4})$ out of the term in braces, and simultaneously multiplying within the braces by the Taylor series for $\exp(- a q^{4})$, following which the $q^{4}$ term vanishes from within the braces.  The lead correction to the exponentials is then $q^{6} ( -2 \overline{ X^{2}}^{3} + \overline{X^{2}}\  \overline{X^{4}} -  2 \overline{X^{6}} )/48$.  Iterating the process of factoring out lead terms as exponentials,
eq \ref{eq:g1defP2a} can be rewritten through ${\cal O}(q^{8})$ as
\begin{displaymath}
 g^{(1)}(q,\tau) =
    \exp\left[- \left(q^2 \frac{\overline{X^{2}}}{2} - q^{4}\frac{
    (\overline{X^{4}} - 3
    \overline{X^{2}}^{2}) 
    }{24}+     q^{6} \frac{( 30 \overline{X^{2}}^{3} - 15   \overline{X^{2}} \  \overline{X^{4}} + \overline{X^{6}} )}{720} \right.  \right.
\end{displaymath}
\begin{equation}
\left. \left. -q^{8} \frac{(630   \overline{X^{2}}^{4}-     420 
\overline{X^{2}}^{2} \overline{X^{4}} + 35  \overline{X^{4}}^{2}
 + 28  \overline{X^{2}} \  \overline{X^{6}}
- \overline{X^{8}} )}{40320} \right) \right] (1 + {\cal O}(q^{10})+\ldots
      \label{eq:g1Pexpanded}
\end{equation}

Equation \ref{eq:g1Pexpanded} is the general expansion for the incoherent structure factor in terms of moments of the distribution of particle displacements, complete through order $q^{8}$.  The mean-square displacement by itself determines only the lead term of $g^{(1)}(q,\tau)$.  

For nearly Brownian particles, $\overline{X^{2n}}$ increases approximately as $t^{n}$.  If $g^{(1)}(q,\tau)$ is not a simple exponential, the terms of eq \ref{eq:g1Pexpanded} of order $q^{4}$ and higher do not vanish, and $\log(g^{(1)}(q,\tau))$ does not give the mean-square particle displacement. It might superficially appear that complications arising from terms of ${\cal O}(q^{4})$ and higher could be avoided by moving to sufficiently small $q$.  Ignoring the modest detail that in conventional units under any credible experimental circumstances $q \gg 1$, the superficial appearances are misleading. To clarify this, we examine a special case.

\section{Spectrum of a Bidisperse System}

Suppose that the field correlation function is exactly
\begin{equation}
    g^{(1)}(q,\tau) = A_{1} \exp(- D_{1} q^{2} \tau) +  A_{2} \exp(- D_{2} q^{2} \tau).
    \label{eq:twocomponent}
\end{equation}
Here $A_{1}$ and $A_{2}$ are two mode amplitudes and $D_{1}$ and $D_{2}$ are two diffusion coefficients.  This $g^{(1)}(q,\tau)$ would arise from a bidisperse suspension of Brownian particles; it would also arise from a monodisperse suspension in which Brownian particles found themselves in two different environments.  Each mode separately corresponds to particles whose motions are describes by a Gaussian Markoff process with $\overline{X^{2}} = 2 D_{i} \tau$.  

Equation \ref{eq:twocomponent} is a uniform function of $q^{2}\tau$.  If one changes the scattering angle, thereby changing $q$, and simultaneously changes the units of time, so that $q^{2} \tau$ does not change (equivalently, if one plots the spectrum against $q^{2} \tau$ rather than against $\tau$), then the shape of the spectrum is \emph{independent} of $q$.  By going to lower scattering angle, one reduces $q$, causing each of the two decay modes to appear at later times, but other than a change by a constant multiplicative factor in the temporal positions of all spectral features, the spectral lineshape does not change when $q$ is changed.

The field correlation function is numerically well-behaved so it and its logarithm have Taylor series in $q^{2}$.  The simple Taylor series is
\begin{equation}
       g^{(1)}(q,\tau) = (A_{1}+A_{2})\left(1 - \frac{(A_{1} D_{1} + A_{2} D_{2})}{(A_{1}+A_{2})} q^{2} \tau  +  
       \frac{(A_{1} D_{1}^{2} + A_{2} D_{2}^{2})}{2(A_{1}+A_{2})}  (q^{2} \tau)^{2} - \ldots)\right.
       \label{eq:twocomponentseries}
\end{equation}

The exponential of the Taylor series of the logarithm of eq \ref{eq:twocomponent} is
\begin{equation}
   g^{(1)}(q,\tau) = (A_{1}+A_{2}) \exp\left( - \frac{(A_{1} D_{1} + A_{2} D_{2})}{(A_{1}+A_{2})} q^{2} \tau  +  
           \frac{A_{1}A_{2} (D_{1} - D_{2})^{2} }{2 (A_{1}+A_{2})^{2} }   (q^{2} \tau) ^{2} + \ldots \right),
   \label{eq:twocomponentlogseries} 
\end{equation}
Note that the $q^{4}$ terms of eqs \ref{eq:twocomponentseries} and \ref{eq:twocomponentlogseries} are not the same. 
Even in our special case, in which the motion of each particle is separately totally characterized by its mean-square displacement as a function of time, the scattering spectrum includes nontrivial terms determined by the difference $(A_{1}A_{2} (D_{1} - D_{2})^{2} )/(2 (A_{1}+A_{2})^{2})   (q^{2} \tau) ^{2}$ between the actual mean-fourth displacement, and the mean-fourth displacement expected for Gaussian particles having the same mean-square displacement. 

The series in eq \ref{eq:twocomponentlogseries} recovers the original function {\em and all of its properties}. In particular,  $g^{(1)}(q,\tau)$ as plotted against $q^{2}\tau$ is invariant under a change in $q$ and, therefore,\emph{ eq  \ref{eq:twocomponentlogseries} as plotted against $q^{2} \tau$ is invariant under a change in $q$.} The relative importance of the $q^{2}$ and high-order terms does not change as $q$ is changed, so eq \ref{eq:twocomponentlogseries} does not tend toward a single exponential in $\tau$ as $q \rightarrow 0$. If $q$ is changed by a factor $f$, the time scale on which eq  \ref{eq:twocomponentlogseries} relaxes changes by a factor $f^{-2}$, but the spectral lineshape does not change.  

In our special case the $A_{i}$ and the $D_{i}$ are simply numbers, not functions of time.  Therefore, comparison with the orthodox spectral cumulant expansion
\begin{equation}
   g^{(1)}(q,\tau) = \exp\left(\sum_{j=0}^{\infty} \frac{K_{j} (-\tau q^{2})^{j}}{j!}\right),
     \label{eq:cumulantseries}
\end{equation}
where the $K_{j}$ are the spectral cumulants, is directly possible.  The series in eqs \ref{eq:twocomponentlogseries} and \ref{eq:cumulantseries} (after taking logarithms) are both convergent power series in $\tau$, so they are equal term by term, leading to
\begin{equation}
    K_{0} = \log(A_{1}+A_{2}) 
\end{equation}
\begin{equation}
    K_{1} = \frac{(A_{1} D_{1} + A_{2} D_{2})}{A_{1}+A_{2}}q^{2}
\end{equation}
\begin{equation}
    K_{2}=  \frac{A_{1}A_{2} (D_{1} - D_{2})^{2} }{2 (A_{1}+A_{2})^{2} }q^{4}
    \label{eq:cumulantvalues}
\end{equation}
The first cumulant, which determines the initial decay of the field correlation function, is a weighted average of both diffusion coefficients; the initial decay rate does not give the diffusion coefficient for the faster mode.
The second cumulant is the $q^{4}$ term.  Except in systems containing highly monodisperse particles in very simple fluids, higher spectral cumulants are not negligible.  Because the spectral lineshape is invariant to a change in $q$ the higher cumulants cannot be made less important by going to smaller $q$.

Equations \ref{eq:cumulantvalues} refer to the model given by eq \ref{eq:twocomponent}.  In general, the cumulants may be obtained from logarithmic derivatives of $g^{(1)}(q,\tau)$, namely for $n \geq 1$
\begin{equation}
     K_{n} =  \lim_{t \rightarrow 0} \left(-\frac{\partial}{\partial t} \right)^{n} \log\left(g^{(1)}(q,\tau)\right).
    \label{eq:Kngeneral}
\end{equation}
In terms of the particle displacements, the first few cumulants are
\begin{equation}
     K_{1} =  \lim_{t \rightarrow 0}\left(  \frac{q^2}{2} \frac{\partial \overline{X^{2}}}{\partial t} 
           -   \frac{ q^{4}}{24}  \frac{\partial(\overline{X^{4}} - 3 \overline{X^{2}}^{2})}{\partial t} 
   +\ldots \right)
    \label{eq:K1general}
\end{equation}
and 
\begin{equation}
     K_{1} =  \lim_{t \rightarrow 0}\left(  \frac{q^2}{2} \frac{\partial^{2} \overline{X^{2}}}{\partial t^{2}} 
           -   \frac{ q^{4}}{24}  \frac{\partial^{2}(\overline{X^{4}} - 3 \overline{X^{2}}^{2})}{\partial t^{2}} 
   +\ldots \right)
    \label{eq:K2general}
\end{equation}
For the Brownian particles of the special case, $\overline{X^{2}} \sim t^{1}$ and $\overline{X^{4}} \sim t^{2}$, so only the $\overline{X^{2}}$ term contributes to $K_{1}$ and only the $\overline{X^{4}}$ term contributes to $K_{2}$. This simplification cannot be made without some specification of the physical properties of the system.  Also, in interpreting the limit $t \rightarrow 0$, the original field correlation function $g^{(1)}(q,\tau)$ is only measured at a series of times $\tau$ that are much larger than the relaxation time $\tau_{b}$ of the diffusing particles, so the limit also only reaches to times $\gg \tau_{B}$. Care must then be taken to interpret the derivatives in the above equations as examining correlations between particle positions at some time $t$ and particle velocities at later times $t+\tau$; without this interpretation anomalous outcomes may follow\cite{phillies1984a}.  

How does taking the $q \rightarrow 0$ limit affect eq \ref{eq:twocomponentlogseries}?  Suppose the limit of small $q$ is taken. In this case, a point with fixed $\tau = \tau_{o}$ refers to a point earlier and earlier in the spectrum's decay.  If $q$ is truly small, $g^{(1)}(q,\tau) = 1$ to within the precision of the experiment.  At slightly larger $q$, the decay of $g^{(1)}(q,\tau)$ will from 0 to $\tau_{o}$ be linear in $\tau$, and will therefore within the precision of the experiment be indistinguishable from a pure exponential having $K_{1}$, above, as its relaxation rate.  Note, however, that this description of $g^{(1)}(q,\tau)$ as a purely $q^{2}$-dependent exponential only refers to very early times.  At later times, the ${\cal O}(q^{4})$ and higher terms of $g^{(1)}(q,\tau)$, whose coefficients are independent of $\tau$ and are therefore present at all times, rise above the limits of experimental precision.  No matter how small the non-zero $q$, if $\tau$ is increased enough that the field correlation function relaxes to zero, all the higher-order-in $q$ terms of eq \ref{eq:twocomponentlogseries} become substantial.

\section{Discussion}

The objective of this short paper was to show how the incoherent structure factor $g^{(1)}(q,\tau)$ measured by light scattering spectroscopy is related to the distribution of particle displacements during the interval $\tau$.
To put the discussion into its correct historical context, the standard result, eq \ref{eq:g1deltar2} for ideal Brownian particles, was obtained both from the Langevin equation and from Doob's Theorem.  For ideal Brownian particles $- \log(g^{(1)}(q,\tau))$ is determined by the mean-square particle displacements $\langle (\Delta \mathbf{R})^{2} \rangle$.  

As seen in eq \ref{eq:g1Pexpanded}, lead terms for the general relationship between $g^{(1)}(q,\tau)$ and the average particle displacements were obtained.  Except at very small times, $g^{(1)}(q,\tau)$ is in general determined not only by the mean-square particle displacements, but also by the higher moments of the particle displacement distribution. Plots of $-\log(g^{(1)}(q,\tau))$ against $t$, taken to long times, do not reveal the mean-square particle displacement through long time, except in the special case of ideal Brownian particles, for which $g^{(1)}(q,\tau)$ is a pure exponential.  However, if the spectrum is an exponential, there is no need to determine a limiting slope at large $t$, because the slope is the same at all times, and therefore the short-time behavior determines $\overline{X^{2}}$ correctly.  There are non-exponential QELSS spectra of probes, in complex fluids, that have already been analyzed by applying eq \ref{eq:g1deltar2}. It is important to emphasize that nothing is wrong with the underlying spectra.  Re-analysis of those spectra using orthodox QELSS approaches\cite{cumulants,streletzky1999a} may well yield interesting information about the fluids. 

There remains the case of particles whose dynamics are complex at short time, but whose motion when adequately coarse-grained appears approximately Brownian.  This case would arise, for example, for Brownian particles passing through an ordered or random potential energy field having a well-defined longest length scale $\xi$.  Particle motions over distances $\gg \xi$ would smooth over the variations in the potential energy field, so that if trapping was not an issue the particle motions over very long distances would be approximately Brownian and eq \ref{eq:g1deltar2} would again be applicable, even though the short-time motions were not Brownian.  

However, there is no guarantee that the dynamics of a complex fluid are appropriately characterized as having a longest length scale, or that the spectrum is a single exponential at long times.  Many complex fluids have hydrodynamic interactions, whose longest range is given by the Oseen tensor with its $r^{-1}$ interaction, leading to relaxation effects on all distance scales.  Correspondingly, relaxations are found with all time scales, leading to a $g^{(1)}(q,\tau)$ having a long-time stretched-exponential-in-time relaxation, exactly as seen experimentally\cite{streletzky1999a} in some cases.  It is, of course, possible to plot the log of a stretched-exponential $g^{(1)}$ against time out to long times, but that plot shows a smooth curve that lacks a long-time straight line limit.  One could always fit the last few points of such a $g^{(1)}$ to a straight line, but that line is only the local tangent of a smooth curve that continues to bend at longer times.  The slope of that fitted tangent line is determined by the largest $\tau$ at which one measured $g^{(1)}$, and is therefore purely an artifact of the fitting process.

In the above, it has been shown that in general QELSS spectra do not determine mean-square particle displacements $\overline{X^{2}}$.  The special-case exception to this rule, in which QELSS spectra do determine $\overline{X^{2}}$, refers to identical particles performing Brownian motion in a simple homogenous fluid.   In this special case, the spectrum is a pure exponential, and all spectral cumulants other than $K_{0}$ and $K_{1}$ vanish. 

How does this result affect interpretation of the literature?  The optical probe diffusion literature, e.g., refs.\ \onlinecite{lin1982a,russo1989a,pu1989a,cao1997a,streletzky2003a}, appears to be unperturbed by the above results, because in the optical probe literature eq.\ \ref{eq:g1deltar2} is not used to interpret QELSS spectra.  Instead, QELSS spectra are fit to exponentials, to cumulant series or sums of cumulant series, or to special functions, or are subject to Laplace inversion.  With any of these fits, the objective is to generate a short list of parameters that characterize the spectra, and to examine how the fitting parameters are related to other solution properties or to fundamental models for polymer solution dynamics. Because eq \ref{eq:g1deltar2} was not used, it is not significant that it is not correct for probes in complex fluids.

A rather different situation is found in the microrheology literature, that is experiments tracing themselves back to Ref.\ \onlinecite{mason1995a}, in which eq \ref{eq:g1deltar2} is systematically applied to interpret light scattering by optical probes.  Much of the microrheology literature uses diffusing wave spectroscopy; the above analysis did not determine if eq \ref{eq:g1deltar2} is correct for DWS measurements.  However, some of the microrheology literature, e.g., Mason, et al.\cite{mason1996a} proposes that QELSS 'can also be used to measure the mean square displacement of probe particles', and therefore can be used in the same way that DWS  can be used, a proposal that as seen above is incorrect.  It is important to emphasize that the difficulty is in applying eq \ref{eq:g1deltar2} to the spectra, that nothing is wrong with the spectra themselves, and therefore that spectra can be correctly reinterpreted using  
methods developed for studies of optical probe diffusion.  As examples of the approach envisioned by Mason, et al.\cite{mason1996a}, consider:

1) Dasgupta, et al.\cite{dasgupta2002a}, studied probe diffusion in polyethylene oxide: water using DWS, QELSS, and eq \ref{eq:g1deltar2}.  Their QELSS data, using very large (0.97 $\mu$m) probes in 4wt\% polyethylene oxide-water, was reduced via eq \ref{eq:g1deltar2} to a time-dependent mean-square displacement, whose angular dependence scales as $q^{2}$.  As noted by Dasgupta, et al., at short times their QELSS and DWS data begin to differ, and their QELSS data was not reported for times shorter than a large (by QELSS standards) 10 mS.  The deviation is what would plausibly be observed if DWS followed eq \ref{eq:g1deltar2} even though QELSS follows eq  \ref{eq:g1Pexpanded}.  Our own\cite{oconnell2005} QELSS data on probes in aqueous 900 kDa polyethylene oxide, extended down to a few $\mu$S, confirms that optical probe spectra in these solutions at short times are radically nonexponential, and therefore that eq \ref{eq:g1deltar2} is not applicable for probes in polyethylene oxide:water as studied by QELSS.

2) van der Gucht, et al.\cite{vandergucht2003a} used QELSS to study diffusion of 250 nm probes in solutions of the associating polymer bis(ethylhexylurido)toluene as a function of polymer concentration.  Their probe spectra resemble those of probes in solutions of hydroxypropylcellulose\cite{streletzky2003a}, probes showing a near-exponential relaxation at small polymer concentration and an additional, prominent, long-lived mode at elevated polymer concentrations.  Here $g^{(1)}(q,\tau)$ is interpreted in terms of $\overline{X^{2}}$, with early and late regions of the spectrum interpreted in terms of diffusive decays (i.e., $\overline{X^{2}} \sim t^{1}$).  

3) Kang, et al.\cite{kang2005a} report on the diffusion of spherical probes through solutions of fd-viruses, which are effectively rigid rods, using fluorescence correlation spectroscopy, video microscopy, and QELSS to measure diffusion of optical probes in different time regimes.  They assume that QELSS gives the mean-square particle displacement, an assumption based on an expansion in terms of small wave vectors, and assert that they are able to enter this regime experimentally.  In this regime, as discussed above, $g^{(1)}(q,\tau)$ is necessarily indistinguishable from a single exponential that has decayed little from its average value.
 
Finally, in searching the literature on this topic, note that the optical probe and microrheology literatures are appreciably non-communicating; a citation search of either literature will largely not couple into the other.

\end{document}